# [#]Non-Perturbative, Unitary Quantum-Particle Scattering Amplitudes from Three-Particle Equations[#]

James Lindesay[a,b] and H. Pierre Noyes[a]

[a]Stanford Linear Accelerator Center, Stanford University, Stanford, California 94309
[b]Computational Physics Laboratory, Howard University, Washington, D.C. 20059

**Abstract**

We here use our non-perturbative, cluster decomposable relativistic scattering formalism to calculate photon-spinor scattering, including the related particle-antiparticle annihilation amplitude. We start from a three-body system in which the unitary pair interactions contain the kinematic possibility of single quantum exchange and the symmetry properties needed to identify and substitute antiparticles for particles. We extract from it unitary two-particle amplitude for quantum-particle scattering. We verify that we have done this correctly by showing that our calculated photon-spinor amplitude reduces in the weak coupling limit to the usual lowest order, manifestly covariant (QED) result with the correct normalization. That we are able to successfully do this directly demonstrates that renormalizability need not be a fundamental requirement for all physically viable models.

## I. INTRODUCTION

In this paper we begin to demonstrate how to extract quanta as boundary states from our non-perturbative, cluster decomposable relativistic scattering formalism[1,2] by concentrating on the extraction of a quantum-particle scattering amplitude from an appropriate three-body system. This system, described by three body integral equations of the Faddeev type for a unitary *T*-matrix, consists of a source , the particle of interest, and a detector. The source cluster is postulated to interact with the particle via a two-body unitary scattering amplitude from which one can extract the quantum in question. In order for this process to include the kinematic possibility of describing the emission of a quantum that can be interpreted as a boundary state, this scattering must be *anelastic* in the sense that the source changes mass when the quantum is emitted. Similarly, the detector must engage in a unitary *anelastic* two-body scattering with the particle. To convince the reader that we have done this properly, we illustrate the process by extracting the scattering of a massless, transverse photon with a spinor and computing Compton scattering, achieving the usual manifestly covariant (QED) result in the weak coupling limit.

---

[#] Work supported in part by Department of Energy contract DE-AC03-76SF00515

In order to achieve this result, we must rely on several properties of our formalism with which the reader cannot be expected to be familiar. The first is that it is *cluster decomposable;* boundary states are *disentangled* from each other in the usual quantum mechanical sense. This allows them to be complex, although self-interacting or ``fully dressed'' (described by physically observable masses, charges,…), and to enter only kinematically into the dynamics described by the integral equations. Using our notation[2], if the spectating cluster or particle has label *a* for the decomposed system, there are no interactions or entanglements between that cluster and the other dynamical particles, which implies that $T_{(d)}=0$ for $d \neq a$. Thus, the dynamical equations for the fully connected and transition amplitudes become

$$W_{ab}(Z) = -\bar{\delta}_{ab}T_{(a)}(Z)R_o(Z)T_{(b)}(Z) - \sum_d \bar{\delta}_{ad}T_{(a)}(Z)R_o(Z)W_{db}(Z) \Rightarrow 0$$

$$T_{ab}(Z) = \delta_{ab}T_{(a)}(Z) + W_{ab}(Z) \Rightarrow \delta_{ab}T_{(a)}(Z)$$

1. 1

The use of the appropriate off-shell parameter $\zeta_{(a)}$ and the off-diagonal parameter $\omega_{(a)}$ for each of the clusters insures that the sub-systems can be appropriately disentangled according to the actual physical scattering process being described, where

$$e_a(M_o u^0_{(a)}) \equiv u^0_{(a)}\sqrt{m_a^2 + M_o^2 u^2_{(a)}} - M_o u^2_{(a)}$$

$$\zeta_{(a)} = \frac{Z - e_a}{u^0_{(a)}}, \qquad \omega_{(a)} = \frac{M - e_a}{u^0_{(a)}}$$

1. 2

The off-shell parameter Z has only parametric dependency on the relativistic kinematics of the spectator and the on-shell invariant energy $M_o$ of the total system. Since the off-diagonal parameter will have this same dependency upon the spectator kinematics, the dynamical spectrum of the pair (including the bound state spectrum) remains unchanged; it is only expressed using total invariant energy kinematics. The factor $u_{(a)}^0$ in the denominators of these parameters is just the appropriate Lorentz factor that gives the invariant cluster energies in the total system center of momentum frame.

The ability to go off shell in a Lorentz invariant way is an advantage our formalism has over other approaches. We can include bound states (including confined states) directly in a unitary formalism which specifies how to extract elastic scattering, rearrangement collisions and breakup in such a way that the unitarity constraints on all open channels are automatically satisfied. Clearly such processes are non-perturbative. Interpretation of the formalism is aided by the fact that we have shown that it reduces unambiguously to the standard Faddeev formalism in the non-relativistic limit [see ref **2**, Sec. IV.B].

The specific postulate we need for the problem at hand is that the source and detector are disentangled from each other, except through the interaction with the particle, as illustrated below:

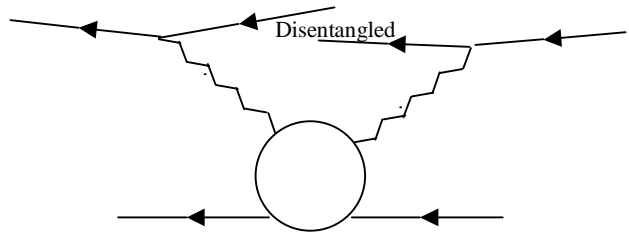

As we will see in this paper, this allows us to separate out the specific process of interest by an appropriate choice of kinematic boundary conditions. The underlying philosophy used here is the Wheeler-Feynman[8] point of view that everything that can be accomplished by treating photons as particles can equivalently be accomplished by treating them as implicit characteristics of particle-particle (and particle-antiparticle) interactions. The equivalence of the two points of view in the case at hand becomes manifest when we show that the amplitude we derive is independent of the structure of the source and the detector, other than that they have the properties already specified above.

Another fact about our formalism that becomes critical in the specific application given here is that we have recently[3] demonstrated that, given any unitary two-particle interaction amplitude, by an appropriate identification and substitution process we can construct both the particle–antiparticle scattering amplitude and the particle-antiparticle transition amplitude with the usual symmetry properties. This includes the explicit dependence of the calculated amplitudes upon *physical* parameters such as mass, charge and other couplings, and on angular momentum *without* a need for renormalizations. In the case of a class of models which reduce to Coulomb scattering in the appropriate limit, we have demonstrated that the expected singularity at a total (virtual) energy of zero is, indeed, present. Consequently, in this paper, we know precisely how to include the appropriate contributions of the anti-particle to the Compton scattering amplitude we derive. This in turn will allow us to embed the related annihilation amplitude in a unitary four body (particle, anti-particle, detector, detector) system and compute the process $p + \bar{p} + D + D \to D^* + D^*$, demonstrating that our formalism is definitely *not* a fixed particle number theory, a fact that is sometimes misunderstood.

In Section II we explain those properties of primary singularities in unitary scattering amplitudes which are made use of in extracting bound states, and which we will make use of in Section III in extracting quanta. Here it is important to realize that when there are bound states present in the complete description of the system, the Faddeev equations necessarily must specify off-shell amplitudes which do not describe physically observable processes. The physically observable (on-shell) amplitudes must then be extracted from them. In general, one must therefore solve the Faddeev equations before extracting some of the amplitudes. Fortunately the process we are considering can be extracted after specifying the appropriate on-shell boundary conditions without this step. It is also important to recognize the distinction between particles that can be created or annihilated only in tandem with the appropriate anti-particles, and quanta that can be singly emitted or absorbed by complex systems. In Section IV we show that the fact that our formalism includes anti-particles allows us to construct the usual causal propagator. In Section V we extract the photon-spinor amplitude, demonstrating explicitly that it is independent of the detailed description of the source and detector of the photons, and has the correct normalization. Section VI presents our conclusions and makes some short comments on new problems that are now open for exploration.

## II.     Extraction of Unitary Amplitudes from Discrete Energies

One of the most useful features of our non-perturbative relativistic few body scattering theory is that, given a unitary (in general off-shell and off-diagonal) T-matrix, we can extract from it unitary physical amplitudes specified by physical values for the masses and charges of the boundary states; these can be represented in a basis describing "free" particles. In particular, when the original T-matrix contains primary singularities in the off-shell energy variable corresponding to bound states in some of the sub-systems, these bound states can be identified as boundary states. This allows one to obtain meaningful expressions for amplitudes that describe systems in kinematic regions for which the usual perturbative expansions cannot provide an analytic handle upon the system's behavior. Scattering theoretic techniques allow the simultaneous description of scattering and bound state behaviors using the same amplitude, which exhibits the analytic structure necessary to describe the system in disjoint energy regimes. In the region of continuum eigenvalues there is usually an overlap in the energy (eigenvalue) spectra of the two descriptions of the system (self-interacting vs. mutual-interacting eigenstates), whereas the discrete eigenvalues are kinematically disjoint (and therefore, eigenvalues of orthogonal states) from the continuum eigenvalues of the mutual-interacting states. We now review how physical amplitudes for processes involving eigenstates of the discrete energies can be extracted from the amplitudes that have been expressed in terms of a complete set of continuum states.

### A.   Primary singularities in few particle amplitudes

The scattering amplitude is typically decomposed into a part corresponding to the identity along with a transition amplitude. For three-particle scattering, several types of transitions are possible. In general, any of the possible boundary states (i.e. three-free or combinations of bound pair + spectator) can exist in either the initial or final state. For the present purpose, we will focus on the elastic scattering and rearrangement amplitudes which result in bound pair + single particle boundary states. For the general case the transition amplitude satisfies

$$<\Psi_\alpha^{(+)}: M, \underline{u} | \Psi_\beta^{(-)}: M_o, \underline{u}_o > = \delta_{\alpha\beta} <\Phi_\alpha: M, \underline{u} | \Phi_\beta: M_o, \underline{u}_o > +$$
$$2\pi i \delta^4(M\vec{u} - M_o\vec{u}_o) A_{\alpha\beta}(\Phi_\alpha | \Phi_\beta: M_o)$$

2. 1

where the other quantum numbers have been temporarily suppressed for conciseness. Here $<\Psi_\alpha^{(+)} | \Psi_\beta^{(-)}>$ is the probability amplitude that is represented by the $S_{\alpha\beta}$ element of the S-matrix, and $\Phi_\alpha, \Phi_\beta$ are (self-interacting) boundary states, which could include the breakup or coalescence channels.

Single particle states are normalized to satisfy

$$<\underline{k}' | \underline{k}> = \frac{\varepsilon(k)}{k_{(s)}^0} \delta^3(\underline{k}' - \underline{k})$$

2. 2

where

$$\varepsilon(k) \equiv \sqrt{m^2 + |\underline{k}|^2}$$

2. 3

The standard state four-vector for a massive particle is written as $\vec{k}_{(s)} = (m,0,0,0)$ whereas for a massless particle it is written as $\vec{k}_{(s)} = (1,0,0,1)$, with the standard momenta having the appropriate units. The corresponding completeness relations can be expressed

$$1 = \int \frac{k_{(s)}^0}{\varepsilon(k)} d^3k \, |\underline{k}><\underline{k}| = \int d^4k \, |\vec{k}><\vec{k}| \, \delta(\sqrt{\vec{k}\cdot\vec{k}} - m).$$

2. 4

This normalization is chosen because amplitudes generated using these states have the usual normalization of momentum states in non-relativistic scattering theory. The scatterings need not preserve particle mass, and generally will not preserve sub-cluster invariant energy. These momentum basis states serve as the basis of most phenomenological data in terms of the asymptotic parameters of the particles. In this sense, the particles described are "fully dressed" (self-interacting) and properly normalized such that the physically measured asymptotic parameters (masses, charges, etc.) are those represented in the states.

The differential cross section will be written as follows using this normalization, assuming the initial state involves only two particles:

$$d\sigma = \frac{(m_{1o}m_{2o})}{\left[(\vec{k}_{1o}\cdot\vec{k}_{2o})^2 - m_{1o}^2 m_{2o}^2\right]^{\frac{1}{2}}} \prod_a \left(\frac{m_a}{\varepsilon_a} d^3 k_a\right)(2\pi)^4 \delta^4\left(\sum_a \vec{k}_a - \vec{k}_{1o} - \vec{k}_{2o}\right)|A_{fo}|^2$$

2. 5

The three-particle transition operator $T(Z)$ can be decomposed using the Faddeev method:

$$T(Z) = \sum_{ab} T_{ab}(Z)$$

2. 6

where the components $T_{ab}$ satisfy

$$T_{ab}(Z) = \delta_{ab} T_{(a)}(Z) - \sum_d \overline{\delta}_{ad} T_{(a)}(Z) R_o(Z) T_{db}(Z)$$

$$\overline{\delta}_{ad} \equiv 1 - \delta_{ad}$$

2. 7

The amplitudes $T_{(a)}(Z)$ represent the two-particle input amplitudes embedded in the three-particle space, and the resolvant $R_o$ is given by

$$R_o(Z) = \frac{1}{H_o - Z}$$

2. 8

where $H_o$ is the Hamiltonian operator whose eigenstates specify the three-free particle continuum. We emphasize the fact that the channel amplitudes $T_{(a)}(Z)$ are unitary in the three-body space, and the resolvant $R_o(Z)$ satisfies the Hilbert identity. Then the transition matrix $T(Z)$ which solves the Faddeev type equation 2. 7 is necessarily unitary[4,5]. Cluster decomposability requires that this resolvant be linear in the energies which add asymptotically.

Diagrammatically the equation for $T_{ab}$ can be represented as follows:

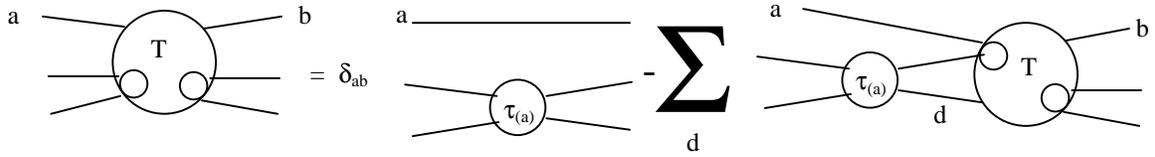

The disconnected term will generate a singular kernel due to products of the delta functions involving the kinematics of the spectator. In our notation, we can define a fully connected amplitude $W_{ab}$ by:

$$W_{ab}(Z) \equiv T_{ab}(Z) - \delta_{ab} T_{(a)}(Z)$$

**2. 9**

Well-defined, non-singular kernels then generate this amplitude. The connected amplitudes $W_{ab}$ satisfy the equation formally expressed by:

$$W_{ab}(Z) = -\bar{\delta}_{ab} T_{(a)}(Z) R_o(Z) T_{(b)}(Z) - \sum_d \bar{\delta}_{ad} T_{(a)}(Z) R_o(Z) W_{db}(Z)$$

**2. 10**

which can be represented by the following diagram:

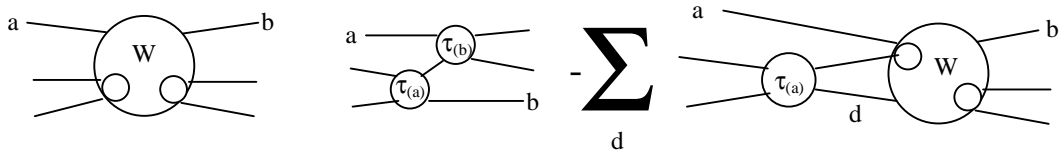

Although the kernels of these amplitudes are non-singular, the amplitudes themselves exhibit singularities in the off-shell parameter Z corresponding to initial and/or final bound state pairs. These "primary singularities" occur at discrete kinematic values corresponding to the expected physical properties of bound boundary states, and will be used to extract the physical amplitudes for those processes.

The full resolvant can be expressed either using the full scattering amplitude $T(Z)$ (which is most conveniently evaluated in terms of the three-free particle continuum basis of states) by:

$$R_F(Z) = R_o(Z) - R_o(Z) \sum_{ab} T_{ab}(Z) R_o(Z)$$

**2. 11**

or in terms of the various channel resolvants[6,7] (which are most conveniently evaluated in terms of the pair-wise mutually-interacting basis of states) by:

$$R_F(Z) = R_o(Z) + \sum_a [R_a(Z) - R_o(Z)] - R_o(Z) \sum_{ab} W_{ab}(Z) R_o(Z)$$

2. 12

We can use this second form of the equation given above to generate the form of the physical amplitudes that represent bound-pair + particle scattering.

The scattering states can be represented in terms of the boundary states using the fully interacting resolvant[1,6,7]

$$\left| \Psi_\alpha^{(\pm)} : M^{(\alpha)}, \underline{u} \right\rangle = \lim_{\eta \to 0} \left[ \mp i\eta R_F(M^{(\alpha)} \pm i\eta) \right] \left| \Phi_\alpha : M^{(\alpha)}, \underline{u} \right\rangle$$

2. 13

where $M^{(\alpha)}$ is the appropriate invariant energy parameter for either a three-free or a pair + spectator boundary state, and $\underline{u}$ is the Lorentz velocity of the system. The fully interacting resolvant includes interactions between the subsystems (sub clusters) of the system. The preservation of the normalization of the basis states expressed in this relationship is guaranteed by the form of the resolvant, which satisfies Hilbert's identity. This means that the amplitudes derived from this resolvant will satisfy unitarity conditions. Thus, the scattering states can be directly extracted from the scattering amplitudes. To do this, it is convenient to define the following set of operations[7]:

$$R_o(Z) W_{ab}(Z) R_o(Z) \equiv R_a(Z) Q_{ab}(Z) R_b(Z)$$

2. 14

These expressions allow the physical transition amplitudes to be directly related to the calculated amplitudes. For pair + spectator elastic scattering and re-arrangement, one utilizes the form of the resolvant given in equation 2. 12 to obtain the result

$$R_F(Z_1) R_F(Z_2) = R_o(Z_1) R_o(Z_2) + \sum_a [R_a(Z_1) R_a(Z_2) - R_o(Z_1) R_o(Z_2)] +$$
$$\sum_{ab} R_a(Z_1) \left\{ \left[ \frac{1}{Z_1 - Z_2} - R_a(Z_2) \right] Q_{ab}(Z_2) - Q_{ab}(Z_1) \left[ \frac{1}{Z_1 - Z_2} - R_b(Z_1) \right] \right\} R_b(Z_2)$$

2. 15

For three-particle to three-particle scattering, the product of the full resolvants takes the form

$$R_F(Z_1)R_F(Z_2) =$$
$$R_o(Z_1)\left\{\left[\frac{1}{Z_1-Z_2}-R_o(Z_2)\right]T(Z_2)-T(Z_1)\left[\frac{1}{Z_1-Z_2}-R_o(Z_1)\right]\right\}R_o(Z_2)$$

2. 16

When going on shell in the parameter Z, the final invariant energy conservation results from the usual relation

$$\frac{1}{M-(M_o\pm i\eta)} \xrightarrow{\eta\to 0^+} \frac{\wp}{M-M_o} \pm i\pi\delta(M-M_o)$$

2. 17

where $\wp$ represents the Cauchy principle part. The action of the resolvants $R_o$ and $R_a$ on the boundary states can be directly determined.

$$\lim_{\eta\to 0}\left[\mp i\eta R_o(M^{(b)}\pm i\eta)\right]\left|\Phi_a:M^{(a)},\underline{u}\right\rangle = 0$$
$$\lim_{\eta\to 0}\left[\mp i\eta R_b(M^{(b)}\pm i\eta)\right]\left|\Phi_a:M^{(a)},\underline{u}\right\rangle = \delta_{ab}\left|\Phi_a:M^{(a)},\underline{u}\right\rangle$$

2. 18

Therefore, by appropriately choosing the off-shell energy parameters $Z_1$ and $Z_2$ in equation 2. 15 we can directly determine the form of the physical amplitude that expresses the overlap of the incoming and outgoing quantum scattering states:

$$<\Psi_a^{(+)}:M,\underline{u}\,|\,\Psi_b^{(-)}:M_o,\underline{u}_o> = \delta_{ab}<\Phi_a:M,\underline{u}\,|\,\Phi_b:M_o,\underline{u}_o>+$$
$$-2\pi i\delta(M-M_o)<\Phi_a:M,\underline{u}\,|\,Q_{ab}^{(+)}(M_o)\,|\,\Phi_b:M_o,\underline{u}_o>$$

2. 19

We next need to demonstrate the extraction of these physical amplitudes from the amplitudes that directly exhibit the primary singularities.

### B. Extraction of singularities from amplitudes

A physical boundary state cluster that includes a bound state as one of its constituent systems will have a "primary singularity" [see references 2,6] in the off-shell parameter Z corresponding to that bound state in the three-particle space. We can use the behavior of the amplitudes in two-particle scattering theory to determine the nature of these primary singularities. The Lippmann-Schwinger equation is most useful for examining the general off-shell behavior of the scattering amplitude; written in operator form it is

$$t_{(a)}(\zeta) = v_{(a)} - v_{(a)} r_{(a)}(\zeta) v_{(a)}$$

**2. 20**

where the interactions are contained in the terms $v_{(a)}$, and $r_{(a)}(\zeta)$ is the mutually interacting resolvant for the pair *(a)*. In terms of the two-particle amplitudes that get properly embedded in the three-particle space, the singularity in the invariant off-shell energy parameter occurs only in the *vrv* term of the previous equation. Therefore, one can "extract" the vertex behavior of the bound states of a scattering pair by inserting a complete set of bound + scattering states for the pair *(a)* in the *vrv* term and evaluating the form in a basis of "free" states. The amplitude $\tau_{(a)}$ is then obtained from the amplitude $t_{(a)}$ (which includes the kinematic delta functions) and satisfies the equation

$$\lim_{\zeta \to \varepsilon_{\mu_{(a)}}} \left( (\zeta - \mu_{(a)}) \tau_{(a)}(M_{(a)}, \hat{q}_{(a)} | M'_{(a)}, \hat{q}'_{(a)}; \zeta) \right) =$$
$$(M_{(a)} - \mu_{(a)}) \psi_{(a)}(M_{(a)}, \mu_{(a)}, l, l_z) Y_l^{l_z}(\hat{q}_{(a)}) Y_l^{l_z*}(\hat{q}'_{(a)}) \psi_{(a)}^*(M'_{(a)}, \mu_{(a)}, l, l_z)(M'_{(a)} - \mu_{(a)})$$

**2. 21**

Here, $M_{(a)}$ is the invariant energy of the pair *(a)*. The wave functions are normalized according to the convention

$$\int dM'_{(a)} \left| \psi_{(a)}(M'_{(a)}; \mu^s_{(a)}, l_{(a)}, l_{z(a)}) \right|^2 = 1$$

**2. 22**

One utilizes this singular behavior in the two-particle amplitudes, which will manifest in the three-particle scattering, to extract the physical amplitudes $A_{ab}$ from the scattering amplitudes $W_{ab}$. Note that the scattering amplitudes *T(Z)* and *W*ab which satisfy the Faddeev equations are *not* themselves the physical amplitudes when there are bound states. The amplitudes for pair + spectator elastic scattering and rearrangement can be calculated using

$$\lim_{Z \to (\varepsilon_{bo} + \varepsilon_{\mu_{(b)o}}) \pm i0} [(Z - (\varepsilon_a + \varepsilon_{\mu_{(a)}}))(Z - (\varepsilon_{bo} + \varepsilon_{\mu_{(b)o}}))] < \underline{k}_1 \underline{k}_2 \underline{k}_3 | W_{ab}(Z) | \underline{k}_{1o} \underline{k}_{2o} \underline{k}_{3o} > \otimes$$

$$\mu_{(a)}^3 \frac{\varepsilon_{\mu_{(a)}}}{\mu_{(a)}} \delta^3(\mu_{(a)} \frac{\underline{k}_{a+} + \underline{k}_{a-}}{M_{(a)}} - \underline{p}_{(a)}) \mu_{(b)}^3 \frac{\varepsilon_{\mu_{(b)0}}}{\mu_{(b)}} \delta^3(\mu_{(b)} \frac{\underline{k}_{b+o} + \underline{k}_{b-o}}{M_{(b)0}} - \underline{p}_{(b)o}) =$$

$$(M - (\varepsilon_a + \varepsilon_{\mu_{(a)}})) < \underline{k}_{a+} \underline{k}_{a-} | \psi_{(a)}(\underline{p}_{(a)}, \mu_{(a)}, l_{(a)}, l_{z(a)}) > \otimes$$

$$< \underline{k}_a, m_a; \psi_{(a)}(\underline{p}_{(a)}, \mu_{(a)}, l_{(a)}, l_{z(a)}) | Q_{ab}^{(\pm)}(\varepsilon_{bo} + \varepsilon_{\mu_{(b)o}}) | \underline{k}_{bo}, m_b; \psi_{(b)}(\underline{p}_{(b)o}, \mu_{(b)}, l_{(b)o}, l_{z(b)o}) > \otimes$$

$$< \psi_{(b)}(\underline{p}_{(b)o}, \mu_{(b)}, l_{(b)o}, l_{z(b)o}) | \underline{k}_{b+o} \underline{k}_{b-o} > (M_o - (\varepsilon_{bo} + \varepsilon_{\mu_{(b)o}}))$$

2. 23

The wave functions are related to the matrix elements using

$$< \underline{k}'_{a+} \underline{k}'_{a-} | \psi_{(a)}(\underline{p}_{(a)}, \mu_{(a)}, l_{(a)}, l_{z(a)}) > \equiv$$

$$\mu_{(a)}^3 \frac{\varepsilon_{\mu_{(a)}}}{\mu_{(a)}} \delta^3(\mu_{(a)} \frac{\underline{k}'_{a+} + \underline{k}'_{a-}}{M'_{(a)}} - \underline{p}_{(a)}) \frac{\psi_{(a)}(M'_{(a)}; \mu_{(a)}, l_{(a)}, l_{z(a)}) Y_{l_{(a)}}^{l_{z(a)}}(\hat{q}'_{(a)})}{[\rho_{(a)}^{(2)}(M'_{(a)}, m_{a+}, m_{a-})]^{\frac{1}{2}}}$$

2. 24

where the normalization dependent factor $\rho_{(a)}^{(2)}$ is satisfied by the Jacobian factor

$$\rho_{(a)}^{(2)}(M_{(a)}, m_{a+}, m_{a-}) = m_{a+} m_{a-} M_{(a)}^2 | \underline{q}_{(a)}(M_{(a)}^2, m_{a+}, m_{a-}) |$$

2. 25

and the internal momentum $q_{(a)}$ satisfies the well known kinematic relation given by

$$| \underline{q}_{(a)}(M_{(a)}^2, m_{a+}, m_{a-}) |^2 \equiv \frac{[M_{(a)}^2 - (m_{a+} + m_{a-})^2][M_{(a)}^2 - (m_{a+} - m_{a-})^2]}{4 M_{(a)}^2}$$

2. 26

The pair momentum delta functions come from the requirement of a well defined Lorentz frame of reference for the pair as described using either basis of states, and it can be expressed in either of the equivalent forms given by

$$\mu_{(a)}^3 \frac{\varepsilon_{\mu_{(a)}}}{\mu_{(a)}} \delta^3(\mu_{(a)} \underline{u}'_{(a)} - \underline{p}_{(a)}) = u_{(a)}^0 \delta^3(\underline{u}'_{(a)} - \frac{\underline{p}_{(a)}}{\mu_{(a)}})$$

2. 27

where

$$\underline{u}'_{(a)} \equiv \frac{\underline{k}'_{a+} + \underline{k}'_{a-}}{M'_{(a)}}$$

2. 28

represents the spatial components of the four-velocity of the pair system. The alternative form can be useful for extracting the kinematics of particles that are massless and have a different standard state component from massive particles.

The operation described above extracts a singularity from both the initial and final states, resulting in the relation

$$\left\langle \Phi_{(a)} : \underline{k}_a, m_a; \psi_{(a)}(\underline{p}_{(a)}, \mu_{(a)}, l_{(a)}, l_{z(a)}) \middle| Q_{ab}^{(+)}(M_o) \middle| \Phi_{(b)} : \underline{k}_{bo}, m_b; \psi_{(b)}(\underline{p}_{(b)o}, \mu_{(b)}, l_{(b)o}, l_{z(b)o}) \right\rangle \otimes$$
$$\delta(M - M_o) =$$
$$- A_{ab}^{(+)}(\underline{k}_a, m_a; \psi_{(a)}(\underline{p}_{(a)}, \mu_{(a)}, l_{(a)}, l_{z(a)}) \mid \underline{k}_{bo}, m_b; \psi_{(b)}(\underline{p}_{(b)o}, \mu_{(b)}, l_{(b)o}, l_{z(b)o}) : M_o) \otimes$$
$$\delta^4(\vec{k}_a + \vec{p}_{(a)} - \vec{k}_{bo} - \vec{p}_{(b)o})$$

2. 29

(as previously noted in equation 2. 19) where $M_o = \varepsilon_{bo} + \varepsilon_{\mu_{(b)o}}$ and $M = \varepsilon_a + \varepsilon_{\mu_{(a)}}$. Similarly (and more directly), the three-three amplitude can be obtained from

$$\left\langle \Phi_{(0)} : \underline{k}_1, m_1; \underline{k}_2, m_2; \underline{k}_3, m_3 \middle| \sum_{ab} T_{ab}^{(+)}(M_o) \middle| \Phi_{(0)} : \underline{k}_{1o}, m_{1o}; \underline{k}_{2o}, m_{2o}; \underline{k}_{3o}, m_{3o} \right\rangle \otimes$$
$$\delta(M - M_o) =$$
$$- A_{00}^{(+)}(\underline{k}_1, m_1; \underline{k}_2, m_2; \underline{k}_3, m_3 \mid \underline{k}_{1o}, m_{1o}; \underline{k}_{2o}, m_{2o}; \underline{k}_{3o}, m_{3o} : M_o) \otimes$$
$$\delta^4(\vec{k}_1 + \vec{k}_2 + \vec{k}_3 - \vec{k}_{1o} - \vec{k}_{2o} - \vec{k}_{3o})$$

2. 30

The physical (non-singular) amplitudes $A_{\alpha\beta}$ can then be directly substituted into equation 2. 5 to calculate the cross section for this particular type of process.

## C. Unitarity of extracted amplitudes

The unitarity of the amplitudes is a direct reflection of the form of the equation which relates the amplitudes to the resolvants, and the condition on the resolvant which will preserve the normalization of the basis states that are generated by the relation 2. 13. The preservation of normalization of basis states is guaranteed by resolvants which satisfy the Hilbert identity; this form continues to be satisfied for linearly additive energies in a multi-particle system. For resolvants which are related by the equation

$$R_F(Z) = R_o(Z) - R_o(Z)T(Z)R_o(Z)$$

2. 31

the Hilbert identity translates into a unitarity condition on the amplitude $T(Z)$, as can be proven by straightforward algebra. Likewise, if the amplitude T(Z) satisfies the appropriate unitarity condition, the resolvant generates an appropriate transformation between the basis states through equation 2. 13 that will preserve the normalization condition.

The unitarity of the fully off-shell (off-shell and off-diagonal) Faddeev equation follows from the unitarity of the two-particle input amplitudes as was previously noted[4,5]. If the two-particle amplitudes have been properly embedded in the three-particle space for our relativistic theory, the form of the proof remains essentially unchanged[2]. Hence, the unitarity of the multi-channel amplitudes is guaranteed as long as the input amplitudes themselves satisfy the same form of a unitarity condition.

The formal unitarity properties of the amplitudes means that similarly the resolvant will satisfy the appropriate unitarity condition. That the extracted amplitudes will similarly generate a unitary resolvant might not at first glance be as obvious. One might expect from equation 2. 12 that the extra terms will change the normalization of the states. However, one is guaranteed formal unitarity in terms of the original amplitudes. In addition, one might note from equation 2. 18 that only the appropriate resolvants will contribute to the extracted unitarity relations. Indeed this is found to be the case when numerically calculating with amplitudes that have been properly extracted using this formalism. See, for example reference 7, where the extracted amplitudes for pair + spectator elastic scattering, rearrangement, and breakup are shown to be unitary, and inelasticity parameters and cross sections are explicitly calculated.

For completeness, the relationship of the scattering amplitudes described above to the S-matrix description of a scattering process will be briefly demonstrated. One is able to formally construct a scattering operator for transitions involving incoming or outgoing boundary states that include bound states by defining the operator $K_{ab}$ that satisfies

$$R_o(Z) W_{ab}(Z) R_o(Z) \equiv R_o(Z) K_{ab}(Z) R_b(Z)$$

2.32

The equation relating fully mutually interacting eigenstates to boundary states can be expressed by

$$\left| \Psi_\alpha^{(\pm)} : M^{(\alpha)}, \underline{u} \right\rangle = \underset{\eta \to 0}{Lim} \left[ \mp i\eta R_F(M^{(\alpha)} \pm i\eta) \right] \left| \Phi_\alpha : M^{(\alpha)}, \underline{u} \right\rangle$$
$$\equiv U_{(a)}^{(\pm)}(M^{(a)}) \left| \Phi_\alpha : M^{(\alpha)}, \underline{u} \right\rangle$$

2.33

The superscript on the invariant energies $M^{(a)}$ indicates that the energy is kinematically disjoint from channels other than *(a)*. In a discussion of unitarity, all channels that are kinematically possible must be included. This means that given sufficient energy, the breakup of the initial pair must be included as a possible outcome, which is identified as channel 0 for convenience.

$$\left| \Psi_0^{(\pm)} : M, \underline{u} \right\rangle \equiv U_{(0)}^{(\pm)}(M) \left| \Phi_0 : M, \underline{u} \right\rangle$$

2.34

The $U_{(\beta)}$ can then be shown to satisfy

$$U_{(0)}^{(\pm)}(M) = 1 - \underset{\eta \to 0^+}{Lim} R_o(M \pm i\eta) \sum_{a,b} T_{ab}(M \pm i\eta)$$
$$U_{(b)}^{(\pm)}(M^{(a)}) = 1 - \underset{\eta \to 0^+}{Lim} R_o(M^{(a)} \pm i\eta) \sum_a K_{ab}(M^{(a)} \pm i\eta)$$

2.35

These operators can be shown to obey the on-shell orthogonality and completeness relations

$$U_{(\alpha)}^{(\pm)\,t} U_{(\beta)}^{(\pm)} = \delta_{\alpha\beta} \mathbf{1}_\alpha$$
$$\sum_{\alpha=0}^{3} U_{(\alpha)}^{(\pm)} U_{(\alpha)}^{(\pm)\,t} = \mathbf{1} - \wp_{discrete}$$

2.36

where $\wp_{discrete}$ is the projector into the discrete spectrum. The scattering operator can be directly identified in terms of these operators:

$$S_{\alpha\beta}(M) = U^{(+)}_{(\alpha)}(M)^{t} \, U^{(-)}_{(\beta)}(M)$$

2. 37

The demonstration of the operations needed to relate the Faddeev-like amplitudes to the scattering amplitudes exemplifies the non-triviality of the direct expression of unitarity in terms of the amplitudes $Q_{ab}$. This is true because of the possible availability of channels and boundary states (if kinematically accessible) that have a very complicated description in terms of these operators. However, all of the various channel operators as defined connect in a well-defined way to the fully mutually interacting resolvant. We feel that this fact shows this resolvant to be the most convenient quantity for the exploration of the unitarity properties of a scattering system.

### III. Extraction of unitary amplitudes for quanta

Quanta are qualitatively distinct from particles in that they are created by sources and absorbed by detectors. In a finite particle number formulation, the particles have a set of conserved quantum numbers that can only be annihilated by the corresponding antiparticles. In such a formulation, quanta are the carriers of the interactions between particles. However, this does not mean that asymptotic quanta cannot have well defined energy-momentum dispersion relationships on shell. The extraction of quantum entities that exhibit all the kinematic and dynamic characteristics expected in the exchange of quanta is the subject of this section.

#### A. Kinematics of photons

We will consider the actual phenomena of photon particle scatterings that include the source and detection aspects of the photons. An asymptotic photon differs from "near zone" electromagnetic interactions primarily in the ability to assign well defined energy-momentum particulate properties to the interaction of a distant source/sink with the dynamic subsystem. After the emission of a photon, the source becomes disentangled from the quantum process that describes the scattering of the quantum. Similarly, prior to the absorption of the final state photon, the detector is usually presumed to be disentangled from the prior quantum operations during the scattering. Using this perspective, an asymptotic quantum-particle interaction can always be described in terms of the two-body anelastic interaction, one body being the photon source and the second being the particle that absorbs the photon. One needs to consistently extract the photonic behavior from this interaction in a way that gives the expected kinematics. Note that we articulate here a specific instance of the usual Wheeler-Feynman interpretation[8] of photons.

One should thus expect to be able to extract a Compton scattering process from a finite-particle number process that includes a composite source for the initial state photon and a composite detector or sink for the final state photon. This of course would represent the actual experimental arrangement for the measurement of such a process. Diagrammatically, the scattering is represented as follows:

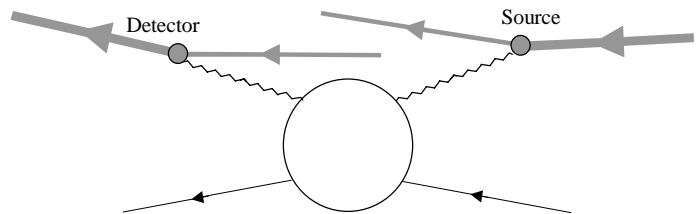

As a three-body problem, an excited source emits a photon, which then scatters from another particle and gets detected at a later time by exciting some detector. The source and the detector are disentangled from each other EXCEPT through the exchange of the photons through the scattering with the third particle. This disallows any direct interaction between the source and the detector. Consequently, in equation 2.10 which describes the dynamics of the three-body system, there are NO contributions from a channel in which the detector and the source scatter directly, nor can there be other interactions between the particle and either the source or detector beyond the single quantum exchange. We immediately conclude that only the driving term contributes to such a scattering process:

$$W_{ab}(Z) = -\bar{\delta}_{ab} T_{(a)}(Z) R_o(Z) T_{(b)}(Z)$$

3.1

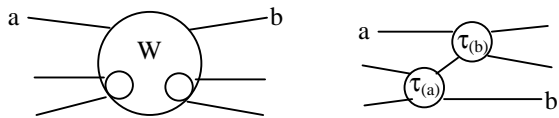

where a and b only refer to the source or the detector. We only need to formally extract the photons from the pair amplitudes, and assure the proper kinematics for particulate photons.

We begin by examining the kinematics of the process. For clarity, a thicker line on the diagram represents an excited source or detector. Kinematics requires that the source mass and detector mass will change during the emission and absorption process in order to conserve both energy and momentum; that is, the two body (source-particle or detector-particle) interactions are *anelastic*. We will need to compare the kinematics of photon scattering while representing photons as asymptotic particles with the kinematics of source-particle-detector scattering. Generally, the center of momentum systems of the different systems need not be the same. Actually, there are numerous different systems that will give equivalent scattering with regards to the Compton process, corresponding to the numerous sources that can produce a given photon. However, the kinematic relationships between the systems are well defined. Examine the following diagrams:

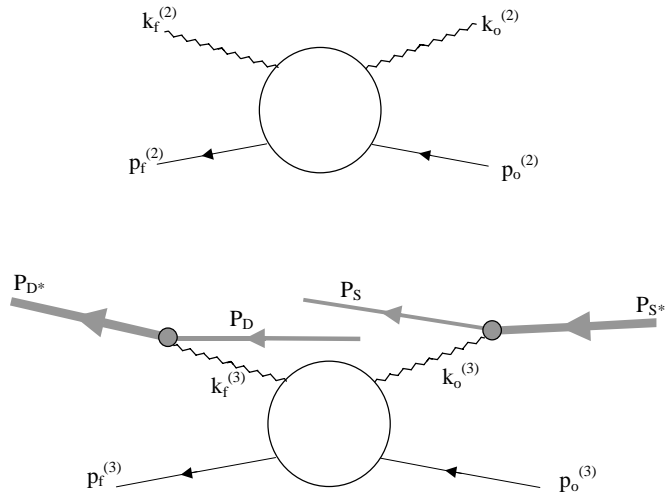

The on-shell kinematics of the systems can be related as follows:

$$\vec{p}_f^{(2)} + \vec{k}_f^{(2)} = \vec{p}_o^{(2)} + \vec{k}_o^{(2)}$$
$$\vec{p}_f^{(3)} + \vec{P}_{D*} + \vec{P}_S = \vec{p}_o^{(3)} + \vec{P}_D + \vec{P}_{S*}$$
$$\vec{P}_{D*} = \vec{P}_D + \vec{k}_f^{(3)} \qquad \vec{P}_{S*} = \vec{P}_S + \vec{k}_o^{(3)}$$

3. 2

The superscripts indicate the number of clusters involved, and are assumed to be written in the center of momentum system which most conveniently describes that particular parameter. The first equation represents energy-momentum conservation in the photon-particle scattering process, the second equation represents energy-momentum conservation in the source-particle-detector scattering process, and the final equations express the relationship between the energy-momenta of the source and detector with the corresponding photons. These equations can immediately be shown to be consistent by direct substitution. Therefore one can ALWAYS describe the process in these kinematically equivalent representations.

We can define a standard frame for which detectors and sources can be found such that $\vec{p}_{f_o}^{(2)} = \vec{p}_{f_o}^{(3)}$, which will imply identical relations for the photon energy-momenta. For such a center of momentum system one can immediately prove that $\underline{P}_S + \underline{P}_D = \underline{0}$. Furthermore, most experimental arrangements have the detector at zero-momentum prior to the scattering. For such arrangements, very straightforward kinematics gives the form of the mass excitation in either the source or the detector necessary to insure asymptotic photon kinematics

$$M^2_{\substack{S^* \\ D^*}} = M_{\substack{S \\ D}}(M_{\substack{S \\ D}} + 2k_{o_f})$$

3. 3

We therefore know that the source and detector must be composite systems whose internal energies can change in a matter that allows the consistent emission or absorption of a photon. Finite particle scattering theory can describe such composite systems in a well-defined manner, and assures the validity of the cluster decomposability that allows us to make the argument just presented.

B. Extraction of photons from amplitudes

Singularities are a necessary aspect of scattering theory that can serve as useful tools or as pesky obstacles. The singularities used to extract bound states in Section II were very useful in obtaining physical amplitudes for scattering processes. The singularities that insure energy or momentum conservation cause some problems when obtaining scattering probabilities from probability amplitudes, particularly when those singularities involve continuous parameters. A standard way to avoid these problems is to discretize the momentum spectrum in a universe of finite volume V, then take the limit as the volume becomes immeasurably large. These discrete box states allow one to formally "square" kinematic delta functions and express the normalizations that follow in terms of hypothetical large volumes which do not appear in the final expression. Any asymptotic boundary state can be handled in this way.

Since we will be extracting asymptotic photons from the two-particle interactions, we will assume that these particles likewise will have a discrete spectrum, which in the final form will be effectively continuous. In the Lippmann-Schwinger equation for the scattering amplitude, this means that the singularity in the resolvant will have a discrete spectrum analogous to the bound state spectra of the amplitudes discussed in Section II. This is equivalent to saying that the source or detector will always undergo a quantized transition in the emission or absorption of an asymptotic photon. Diagrammatically, an emission process will look as follows:

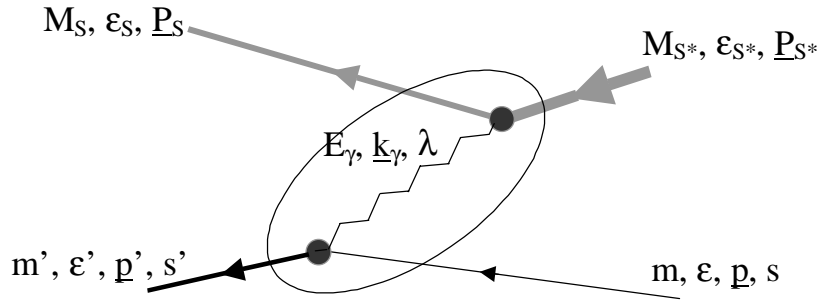

The on-shell kinematics to be extracted will satisfy

$$E_\gamma = \varepsilon_{S^*} - \varepsilon_S \quad , \quad \underline{k}_\gamma = \underline{P}_{S^*} - \underline{P}_S \quad , \quad E_\gamma = |\underline{k}_\gamma|$$

**3. 4**

We expect the amplitude describing the scattering to still be described by the Lippmann-Schwinger equation, written in the form

$$t(z) = \Delta h - \Delta h \, r_{(\gamma)}(z) \, \Delta h$$

**3. 5**

The form $\Delta h$ represents the difference in Hamiltonian operator forms that have asymptotically equivalent boundary states in scattering theory. The first term on the right hand side of this equation is non-singular in the parameter z (in fact has no z dependence), so that only the photon channel resolvant $r_\gamma(z)$ will have the appropriate kinematic singularity corresponding to the photon if the operator expression is evaluated using states that carry the quantum numbers of the photons. Examining the form expected for the intermediate resolvant, we conclude

$$r_{(\gamma)}(\varepsilon_{S*} + \varepsilon \pm i0) \doteq \frac{1}{\varepsilon_S + E_\gamma + \varepsilon - \varepsilon_{S*} - \varepsilon \mp i0} = \frac{1}{E_\gamma - (\varepsilon_{S*} - \varepsilon_S) \mp i0}$$

3. 6

This resolvant will give a delta function that will insure that the on-shell photon energy is appropriately generated by the source. The very meaning of the extraction of an asymptotic photon requires the on-shell condition at the source given by $\vec{k}_\gamma = \vec{P}_{S*} - \vec{P}_S$ as given in equation 3. 4. Cluster decomposability requires that the amplitude be generated by states in the same Lorentz frame[1,2] Since the asymptotic photon carries well defined 4-momentum in the intermediate state, one is able to use this on-shell kinematics at the source to infer that 4-momentum conservation also occurs at the photon-particle vertex

$$\vec{p}' = \vec{k}_\gamma + \vec{p} \quad \Rightarrow \quad \vec{P}_S + \vec{p}' = \vec{P}_{S*} + \vec{p}.$$

3. 7

Using a momentum basis representation for the transition operator, we conclude

$$\underset{z \to E_\gamma + \varepsilon_S + \varepsilon \pm i0}{\text{Lim}} \langle \vec{P}_S, \vec{p}' | t(z) | \vec{P}_{S*}, \vec{p} \rangle \doteq \pm 2\pi i \, \delta(E_\gamma + \varepsilon_S - \varepsilon_{S*}) \langle \vec{p}' | \Delta h | \vec{k}_\gamma, \vec{p} \rangle \langle \vec{k}_\gamma, \vec{P}_S | \Delta h | \vec{P}_{S*} \rangle$$

3. 8

where the intermediate states correspond to fully interacting basis states, and the external states are boundary states. The limiting form picks out the particular photon state that can appear with asymptotic energy $E_\gamma$. It therefore only remains to specify the form of $\Delta h$ in the vertex functions.

Given a Lagrangian for an interacting system, one can generally construct currents from the generators $G_s$ of internal local gauge symmetries from forms such as in the references[9]

$$J_s^\mu \equiv \frac{\partial L}{\partial (D_\mu \psi)} G_s \psi$$

3. 9

These currents satisfy constraints from Noether[10] conditions resulting from the symmetries. For the abelian symmetry that generates quantum electrodynamics, the current defines a conserved charge. These approaches lead quite generally to interactions of the form

$$\Delta h = J^\mu A_\mu$$

3. 10

where of course $A_\mu$ represents the photon field. In particular, one can directly evaluate the form of the vertex function for a Dirac particle using

$$\mathbf{h} = \gamma^0 \left[ \underline{\gamma} \cdot \underline{p} + m\mathbf{1} + \frac{q}{c}\gamma^\mu A_\mu \right] \quad \Rightarrow \quad \Delta\mathbf{h} = \gamma^0 \left[ \frac{q}{c}\gamma^\mu A_\mu \right]$$

3. 11

We normalize the photon field as

$$\hat{A}_\mu(\vec{x}) = \int \frac{d^3k}{2\pi |\underline{k}|} \sum_{\lambda=-1}^{+1} e_\mu(\underline{k},\lambda) \left[ \hat{a}(\underline{k},\lambda) e^{-i\vec{k}\cdot\vec{x}} + \hat{a}^+(\underline{k},\lambda) e^{i\vec{k}\cdot\vec{x}} \right]$$

3. 12

where the polarization vectors satisfy

$$\vec{e}^*(\underline{k},\lambda') \cdot \vec{e}(\underline{k},\lambda) = \lambda^2 \delta_{\lambda'\lambda}$$

$$\sum_{\lambda=-1,0,+1} e_\mu(\underline{k},\lambda) e_\nu^*(\underline{k},\lambda) \equiv \sum_{\lambda=-1,0,+1} \Delta_{\mu\nu}(\underline{k},\lambda) = \eta_{\mu\nu} - \frac{u_\mu k_\nu + u_\nu k_\mu}{\vec{u}\cdot\vec{k}}$$

3. 13

The asymptotic photon emitted by the source will be the same photon that scatters off of the fermion. For on-shell kinematics, the mass-velocity form for a cluster decomposable formulation is equivalent to the energy-momentum form. This will give a form for the transition amplitude embedded in a three-particle space as follows:

$$\frac{\delta(M_f - M_o)u_f^0\delta^3(\underline{u}_f - \underline{u}_o)}{M_f^3} = \delta(E_f - E_o)\delta^3(\underline{P}_f - \underline{P}_o)$$

$$\lim_{Z \to E_\gamma + \varepsilon_S + \varepsilon + \varepsilon_D \pm i0} \langle \vec{P'}_S, \sigma'_S; \vec{p'}, s'; \vec{P'}_D, \sigma'_D | T_{(D)}(Z) | \vec{P}_{S^*}, \sigma_S; \vec{p}, s; \vec{P}_D, \sigma_D \rangle =$$

$$\pm 2\pi i \delta^4(\vec{p'}-\vec{p}-\vec{k}_\gamma)\frac{u^0\delta^3(\underline{u'}-\underline{u})}{M^3}\delta_{\sigma'_D,\sigma_D}\frac{\Gamma^{\mu^*}(\vec{p'},s';\vec{p},s)}{2\pi i}\Delta_{\mu\nu}(\vec{k}_\gamma,\lambda)\frac{\Gamma^\nu_S(\vec{P'}_S,\sigma'_S;\vec{P}_{S^*},\sigma_{S^*})}{2\pi i} \overset{\bullet}{=}$$

$$\pm 2\pi i \delta^4(\vec{p'}-\vec{p}-\vec{k}_\gamma)\delta^3(\underline{P'}_D-\underline{P}_D)\delta_{\sigma'_D,\sigma_D}\frac{\Gamma^{\mu^*}(\vec{p'},s';\vec{p},s)}{2\pi i}\Delta_{\mu\nu}(\vec{k}_\gamma,\lambda)\frac{\Gamma^\nu_S(\vec{P'}_S,\sigma'_S;\vec{P}_{S^*},\sigma_{S^*})}{2\pi i}$$

$$\text{where} \quad \vec{k}_\gamma = \vec{P}_{S^*} - \vec{P'}_S$$

3. 14

and the internal quantum numbers of the source are assumed to change in a way appropriate to create the requisite photon with well defined kinematics and polarization state. The parameters M and $\underline{u}$ represent the three-particle cluster invariant energy and Lorentz 4-velocity components, and the form is chosen to preserve the Lorentz frames of the associated processes. The momentum conserving delta function for the detector momentum comes from the three-particle identification $\underline{P'}_D+\underline{p'}+\underline{P'}_S=\underline{P}_D+\underline{p}+\underline{P}_{S^*}$ and utilizing the source-particle cluster conservation condition.

### C. Unitarity of photons as boundary states

The three body (source-particle-detector) system we examine in this paper is described by a unitary T-matrix because, according to our hypothesis, the two anelastic two body interactions (source-particle and particle-detector) used in writing down the three body equations are unitary and the source-detector interaction does not enter because, again by the same hypothesis, we have used the cluster-decomposability of our formalism to remove this third interaction from our specific problem. As shown above, this leaves us with only the driving term for the connected amplitude $W_{ab}$ from which we extract our photon-particle amplitude in the same way we extracted bound states in Section II. In general, when we extract bound states from a three body system to obtain an elastic scattering amplitude, the unitarity constraint on this amplitude will involve a sum over all the other relevant amplitudes (i.e. rearrangement amplitudes to other bound states and the breakup amplitude) as we discuss in more detail elsewhere[2,3]. However, because of the way we have set up our extraction, the photon-particle scattering amplitude refers only to a scattering in which a single photon and a single particle of specified momenta and energies scatter to a final state containing again only a single photon and a single particle of specified momenta and energies; all other possibilities are excluded. The unitarity condition on this amplitude is automatically satisfied if we apply it in laboratory situations in which these conditions are met. Of course, in the usual laboratory situation where photon-particle scattering is studied, other channels can indeed be open, such as bremstrahlung, multiple photon emission, etc., and we will require a longer calculation to deal with them. It is important to understand that we do NOT attempt these more complicated calculations here. Nevertheless, we believe it significant that we can show that our formalism does allow us to extract the usual result for Compton scattering from a ``free'' electron using our formalism in the weak-coupling limit. We believe that this fact allows us to say that we have shown that our formalism does allow us to extract photons as boundary states from appropriately constructed source-particle and detector-particle interactions.

## IV. Causal Propagator for Three-Particle Amplitudes

We plan to utilize the formalism that has been developed for relativistic cluster decomposable three-body scattering theory[2,3]. When properly formulated, models in such a formalism will be unitary, and one should be able to properly separate components in ways that are appropriate to the particular scattering arrangement being examined.

### A. Disentanglement of source and detector

We next examine the embedding of the amplitudes discussed in Section III into a larger number space. We have already required that the source and detector have no interaction other than through the exchange of the quanta with the particle, which of course is the usual requirement of an actual experiment. We diagrammatically represent the process as follows:

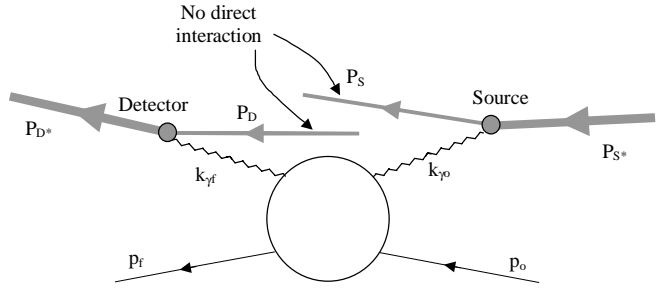

Source and detector are represented again using capital letters for kinematics, with an asterisk representing an excited state. A crucial feature of the Faddeev decomposition is represented in the quantity $\bar{\delta}_{ab} \equiv 1 - \delta_{ab}$. This factor insures all "self-energies" have already been included in the amplitudes $T_{(a)}$, and that any sub-cluster must interact with another sub-cluster before another "self"-interaction. These interactions will of course quantum entangle the various clusters. One can see from the form of the equation 3. 1 for the connected amplitudes $W_{ab}$ that if the source and detector are to remain disentangled except through the exchange of an asymptotic quantum each with the particle, only the driving terms can contribute to that scattering.

$$W_{ab}^{\substack{Driving \\ term}}(Z) = -\bar{\delta}_{ab} T_{(a)}(Z) R_o(Z) T_{(b)}(Z)$$

4. 1

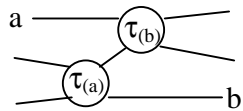

This considerably simplifies the calculation, since any terms involved in the integration of the full relativistic three-particle equation would require a direct interaction between source and detector. We therefore examine this term to determine the scattering amplitude for the process.

The overall amplitude for the scattering process will then be given in terms of the sum over all Faddeev channels given by equation 2. 6. In this case, the identity term corresponds just to the source emitting a photon which is detected by the detector without interacting with the particle, which will be the disconnected term in the channel labeled by the particle as a spectator (i.e. the source-detector channel). The scattering term is then given by the sum over possible channels a and b which result in the source (S) and detector (D) each interacting with the particle (P) only through the exchange of a single quantum. This excludes the possibility of a term $W_{PP}$. Thus

$$T = \delta_{PP} T_{(P)} + W_{DS} + W_{SD}$$

4. 2

The $T_{(P)}$ scattering amplitude corresponds to a single photon emitted by the source and absorbed by the detector. Our formalism requires us to describe this as a component of a unitary scattering amplitude. This term provides the normalization (or calibration) that allows us to calculate the cross section for quantum-particle scattering implied by our extracted amplitude. We find it interesting that our use of the Wheeler-Feynman interpretation forces us to include more of the external experimental setup and calibration in our formalism than is customary in discussions of scattering theory. We will examine the contribution of these terms to the scattering process in section V. A.

### B. Particle-antiparticle correspondences and positive energy projection

The intermediate state in the amplitude will be integrated over all energies for the particles involved. Since the source and detector are asymptotic, and all kinematic information is carried through the photons, these systems will remain as spectators during the dynamical process. We first examine the amplitude $W_{DS}$. The integration over negative particle energies will be associated with positive energy antiparticles with the corresponding momenta using the usual Feynman identifications:

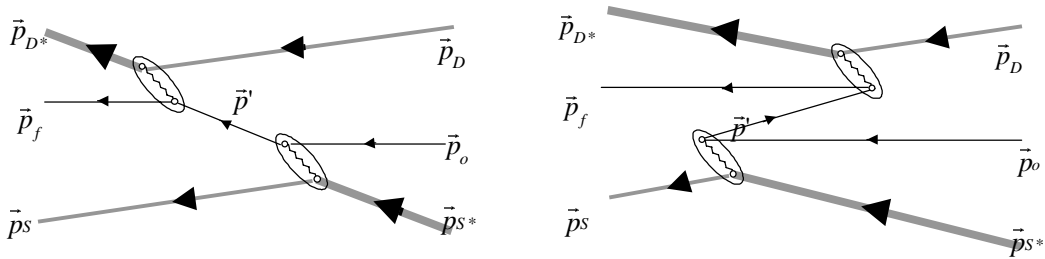

For this process, the resolvant is given by

$$R_o\big((\varepsilon_f + \varepsilon_{D^*} + \varepsilon_S) = (\varepsilon_o + \varepsilon_D + \varepsilon_{S^*}) \pm i0\big) =$$

$$\frac{\wp_+(\vec{p}')}{(\sqrt{p'^2+m^2} + \varepsilon_D + \varepsilon_S) - (\varepsilon_f + \varepsilon_{D^*} + \varepsilon_S) \mp i0} +$$

$$\frac{\wp_+(\vec{p}')}{(\sqrt{p'^2+m^2} + \varepsilon_f + \varepsilon_o + \varepsilon_{D^*} + \varepsilon_{S^*}) - (\varepsilon_o + \varepsilon_D + \varepsilon_{S^*}) \mp i0} =$$

$$\frac{\wp_+(\vec{p}')}{\sqrt{p'^2+m^2} - (\varepsilon_{D^*} + \varepsilon_f - \varepsilon_D) \mp i0} + \frac{\wp_+(\vec{p}')}{\sqrt{p'^2+m^2} + (\varepsilon_{D^*} + \varepsilon_f - \varepsilon_D) \mp i0} =$$

$$\frac{2\sqrt{p'^2+m^2}\;\wp_+(\vec{p}')}{p'^2+m^2 - (E_{\gamma o} + \varepsilon_f)^2 \mp i0}$$

<div align="right">4. 3</div>

where the factor $\wp_+$ represents the projector to positive energies. Including the required kinematics at the particle-final photon vertex, the resolvant can effectively be written in the form

$$R_o\big((\varepsilon_f + \varepsilon_{D^*} + \varepsilon_S) = (\varepsilon_o + \varepsilon_D + \varepsilon_{S^*}) \pm i0\big) \stackrel{\bullet}{=} \frac{2\sqrt{p'^2+m^2}\;\wp_+(\vec{p}')}{m^2 - (\vec{p}')^2 \mp i0}$$

<div align="right">4. 4</div>

For self-interacting Dirac particles, the positive energy projector takes the form

$$\wp_+(\vec{p}') = \left(\frac{\mathbf{p}' + m\mathbf{1}}{2m}\right)$$

<div align="right">4. 5</div>

Combining these terms, one obtains the usual Feynman causal fermion propagator

$$\mathbf{R}_o\big((\varepsilon_f + \varepsilon_{D^*} + \varepsilon_S) = (\varepsilon_o + \varepsilon_D + \varepsilon_{S^*}) \pm i0\big) \stackrel{\bullet}{=} \frac{\sqrt{p'^2+m^2}}{m} \frac{1}{\mathbf{p}' - m\mathbf{1} \pm i0}$$

<div align="right">4. 6</div>

The first factor exactly matches the invariant normalization factor chosen for our (particle) momentum eigenstates.

The examination of the amplitude W$_{SD}$ will involve the analogous diagram:

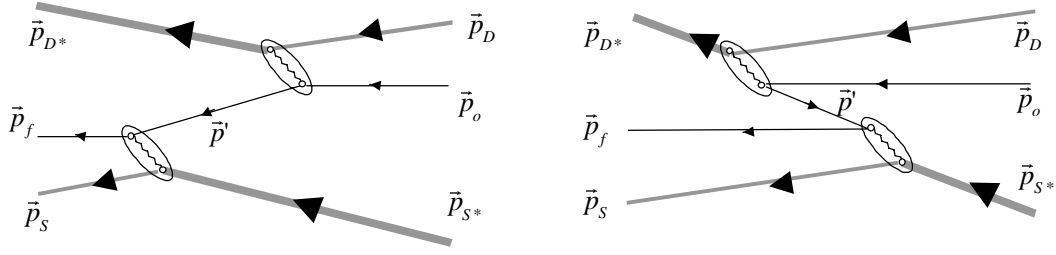

A direct analysis of the intermediate resolvant gives the same form as equation 4.4, with the appropriate kinematics for the intermediate fermion. Of course, the mass parameter in these equations corresponds to the physical mass of the boundary state particle. We therefore will utilize this form of the resolvant when evaluating the physical amplitude that extracts asymptotic quanta in the calculation of the Compton scattering amplitude in the next section.

## V. Photon scattering from spinor particles

By combining the results of the previous sections, we can obtain a form for non-perturbative extracted amplitudes for Compton scattering, if non-perturbative forms for the vertex functions are extracted from the source-charge scattering amplitude. Since the physical amplitudes are extracted from unitary finite particle number amplitudes, they can be unitarily included in multi-channel amplitudes for a given scattering process in a manner outlined elsewhere[2,3,6]. All parameters that will appear in these amplitudes will be well defined in terms of those physically measured for the asymptotic (self-interacting only) boundary states.

### A. Unitary, non-perturbative form

Extracting the asymptotic photons from the calculated amplitude involves examining the limiting behavior of the kinematics required to produce those photons. To explore that behavior and the kinematics of the process, examine the following diagrammatic representation of the process:

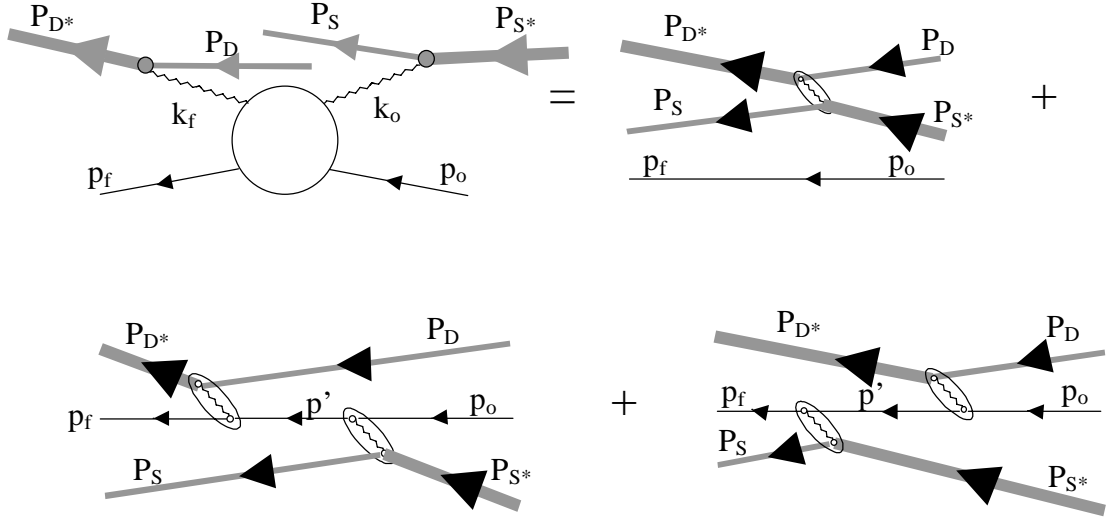

The first term on the right hand side of the above diagram represents a scattering in which the fermion is uninvolved with the photon, which is simply emitted by the source and absorbed by the detector. After being multiplied by the overall invariant energy conserving delta function that is contributed by the resolvants $R_o$ in the amplitude, this term gives a contribution of the form

$$2\pi i\, \delta(E_f - E_o) T_{(P)} \Rightarrow I_P(\vec{P}_{D^*}, \sigma_{z^*}; \vec{p}_f, \vec{P}_S, \sigma_S \mid \vec{P}_D, \sigma_D; \vec{p}_o, s_o;, \vec{P}_{S^*}, \sigma_{S^*}) \equiv$$

$$2\pi i\, \delta_{s_f, s_o} \delta^4\!\left((\vec{P}_{D^*} + \vec{p}_f + \vec{P}_S) - (\vec{P}_D + \vec{p}_o + \vec{P}_{S^*})\right) \delta^4\!\left((\vec{P}_{D^*} + \vec{P}_S) - (\vec{P}_D + \vec{P}_{S^*})\right) \delta_{s_f, s_o} \otimes$$

$$\frac{\Gamma_D^{\mu\,*}(\vec{P}_D, \sigma_D; \vec{P}_{D^*}, \sigma_{D^*})}{2\pi i} \Delta_{\mu\nu}(\vec{k}_\gamma, \lambda) \frac{\Gamma_S^\nu(\vec{P}_S, \sigma_S; \vec{P}_{S^*}, \sigma_{S^*})}{2\pi i}$$

**5. 1**

The next term in the diagram is a fully connected term, corresponding to $W_{SD}$. After being multiplied by the overall invariant energy conserving delta function that is contributed by the resolvants $R_o$ in the amplitude, this term gives a contribution of the form

$$2\pi i \delta(E_f - E_o)T_{(S)}R_o T_{(D)} \Rightarrow$$

$$-(2\pi i)^3 \delta^4\big((\vec{P}_{D^*} + \vec{p}_f + \vec{P}_S) - (\vec{P}_D + \vec{p}_o + \vec{P}_{S^*})\big)A_{SD}(\vec{P}_{D^*},\vec{p}_f,\vec{P}_S \mid \vec{P}_D,\vec{p}_o,\vec{P}_{S^*}) \equiv$$

$$(2\pi i)^3 \delta^4\big((\vec{P}_{D^*} + \vec{p}_f + \vec{P}_S) - (\vec{P}_D + \vec{p}_o + \vec{P}_{S^*})\big)\sum_{s'} \iiint d^4 P'_S \frac{m d^3 p'}{\sqrt{m^2 + p'^2}} d^4 P'_D \frac{u^0 \delta^3(\underline{u}'-\underline{u})}{M^3} \otimes$$

$$\delta^4\big((\vec{P}_{D^*} + \vec{p}_f) - (\vec{P}'_D + \vec{p}')\big)\frac{\Gamma_D^{\mu*}(\vec{P}'_D,\sigma'_D;\vec{P}_{D^*},\sigma_{D^*})}{2\pi i}\Delta_{\mu\nu}(\vec{k}_f,\lambda_f)\frac{\Gamma^\nu(\vec{p}_f,s_f;\vec{p}',s')}{2\pi i}$$

$$\frac{2\sqrt{m^2 + p'^2}}{m^2 - \vec{p}'\cdot\vec{p}'-i0}\wp_+(\vec{p}')$$

$$\delta^4\big((\vec{P}'_S + \vec{p}') - (\vec{P}_{S^*} + \vec{p}_o)\big)\frac{\Gamma^{\alpha*}(\vec{p}_o,s_o;\vec{p}',s')}{2\pi i}\Delta_{\alpha\beta}(\vec{k}_o,\lambda_o)\frac{\Gamma_S^\beta(\vec{P}'_S,\sigma'_S;\vec{P}_{S^*},\sigma_{S^*})}{2\pi i}$$

5.2

The required kinematics that can satisfy the various delta functions is given below:

$$\vec{k}_f = \vec{P}_{D^*} - \vec{P}_D \qquad \vec{k}_o = \vec{P}_{S^*} - \vec{P}_S$$
$$\vec{P}'_S = \vec{P}_S + \vec{k}_f + \vec{p}_f - \vec{p}' \qquad \vec{P}'_D = \vec{P}_D + \vec{k}_o + \vec{p}_o - \vec{p}'$$

5.3

Evaluating the required intermediate state integrals, we obtain

$$A_{SD}(\vec{P}_D + \vec{k}_f, \vec{p}_f, \vec{P}_S \mid \vec{P}_D, \vec{p}_o, \vec{P}_S + \vec{k}_o) =$$

$$\sum_{s'} \frac{\Gamma_D^{\mu*}(\vec{P}_D,\sigma_D;\vec{P}_D + \vec{k}_f,\sigma_{D^*})}{2\pi i}\Delta_{\mu\nu}(\vec{k}_f,\lambda_f)\frac{\Gamma^\nu(\vec{p}_f,s_f;\vec{p}_o + \vec{k}_o,s')}{2\pi i} \otimes$$

$$\frac{\wp_+(\vec{p}_o + \vec{k}_o)}{m^2 - (\vec{p}_o + \vec{k}_o)^2 - i0} \otimes$$

$$\frac{\Gamma^{\alpha*}(\vec{p}_o,s_o;\vec{p}_o + \vec{k}_o,s')}{2\pi i}\Delta_{\alpha\beta}(\vec{k}_o,\lambda_o)\frac{\Gamma_S^\beta(\vec{P}_S,\sigma_S;\vec{P}_S + \vec{k}_o,\sigma_{S^*})}{2\pi i}$$

5.4

Analogously, we can evaluate the amplitude generated by the final term in the diagram

$$A_{DS}(\vec{P}_D + \vec{k}_f, \vec{p}_f, \vec{P}_S | \vec{P}_D, \vec{p}_o, \vec{P}_S + \vec{k}_o) =$$

$$\sum_{s'} \frac{\Gamma^{\alpha *}(\vec{p}_o, s_o; \vec{p}_o - \vec{k}_f, s')}{2\pi i} \Delta_{\alpha\beta}(\vec{k}_o, \lambda_o) \frac{\Gamma_S^{\beta}(\vec{P}_S, \sigma_S; \vec{P}_S + \vec{k}_o, \sigma_{S*})}{2\pi i} \otimes$$

$$\frac{\wp_+(\vec{p}_o - \vec{k}_f)}{m^2 - (\vec{p}_o - \vec{k}_f)^2 - i0} \otimes$$

$$\frac{\Gamma_D^{\mu *}(\vec{P}_D, \sigma_D; \vec{P}_D + \vec{k}_f, \sigma_{D*})}{2\pi i} \Delta_{\mu\nu}(\vec{k}_f, \lambda_f) \frac{\Gamma^{\nu}(\vec{p}_f, s_f; \vec{p}_o - \vec{k}_f, s')}{2\pi i}$$

5.5

The scattering amplitude is obtained from the product of full resolvants after taking an appropriate limit. For the present case, this product is given by equation 2.16, which allows us to express the amplitude as

$$\lim_{\zeta_1 \to 0^+} \lim_{\zeta_2 \to 0^+} [(i\zeta_1)(-i\zeta_2)] \otimes$$

$$< \vec{P}_D + \vec{k}_f, \vec{p}_f, \vec{P}_S | R_F(\varepsilon_D + k_f + \varepsilon_f + \varepsilon_S - i\zeta_2) R_F(\varepsilon_D + \varepsilon_o + \varepsilon_S + k_f + i\zeta_1) | \vec{P}_D, \vec{p}_o, \vec{P}_S + \vec{k}_o > =$$

$$2\pi i [\, I_P(\vec{P}_D + \vec{k}_f, \vec{p}_f, \vec{P}_S | \vec{P}_D, \vec{p}_o, \vec{P}_S + \vec{k}_o) + 2\pi i \delta^4((\vec{k}_f + \vec{p}_f) - (\vec{k}_o + \vec{p}_o)) \otimes$$

$$\{A_{SD}(\vec{P}_D + \vec{k}_f, \vec{p}_f, \vec{P}_S | \vec{P}_D, \vec{p}_o, \vec{P}_S + \vec{k}_o) + A_{DS}(\vec{P}_D + \vec{k}_f, \vec{p}_f, \vec{P}_S | \vec{P}_D, \vec{p}_o, \vec{P}_S + \vec{k}_o)\} \,]$$

5.6

Substituting the forms of the various contributing terms, we obtain

$$2\pi i \delta_{s_f, s_o} \delta^4((\vec{P}_{D*} + \vec{p}_f + \vec{P}_S) - (\vec{P}_D + \vec{p}_o + \vec{P}_{S*})) \otimes$$

$$\Gamma_D^{\mu *}(\vec{P}_D, \sigma_D; \vec{P}_{D*}, \sigma_{D*}) e_{\mu}^*(\vec{k}_f, \lambda_f) \Bigg\{ \delta^4((\vec{P}_{D*} + \vec{P}_S) - (\vec{P}_D + \vec{P}_{S*})) \delta_{s_f, s_o} +$$

$$\sum_{s'} \left(\frac{1}{2\pi i}\right)^2 \Bigg( \Gamma^{\nu *}(\vec{p}_f, s_f; \vec{p}_o + \vec{k}_o, s') e_v^*(\vec{k}_f, \lambda_f) \frac{2m \wp_+(\vec{p}_o + \vec{k}_o)}{m^2 - (\vec{p}_o + \vec{k}_o)^2} e_{\alpha}(\vec{k}_o, \lambda_o) \Gamma^{\alpha}(\vec{p}_o, s_o; \vec{p}_o + \vec{k}_o, s') +$$

$$\Gamma^{\alpha}(\vec{p}_o, s_o; \vec{p}_o - \vec{k}_f, s') e_{\alpha}(\vec{k}_o) \frac{2m \wp_+(\vec{p}_o - \vec{k}_f)}{m^2 - (\vec{p}_o - \vec{k}_f)^2} e_v^*(\vec{k}_f) \Gamma^{\nu *}(\vec{p}_f, s_f; \vec{p}_o - \vec{k}_f, s') \Bigg) \Bigg\} \otimes$$

$$e_{\beta}(\vec{k}_o, \lambda_o) \Gamma_S^{\beta}(\vec{P}_S, \sigma_S; \vec{P}_{S*}, \sigma_{S*})$$

5.7

By dividing out the factors relating to the interaction of the photons with the source and detector, we can extract a photon-fermion amplitude A that can be written in terms of these kinematic parameters only:

$$A(\vec{k}_f, \lambda_f; \vec{p}_f, s_f | \vec{k}_o, \lambda_o; \vec{p}_o, s_o) =$$

$$\sum_{s'} \left(\frac{1}{2\pi i}\right)^2 \Big( \Gamma^{\mu *}(\vec{p}_f, s_f; \vec{p}_o + \vec{k}_o, s') e_\mu^*(\vec{k}_f, \lambda_f) \frac{2m \wp_+(\vec{p}_o + \vec{k}_o)}{m^2 - (\vec{p}_o + \vec{k}_o)^2} e_\nu(\vec{k}_o, \lambda_o) \Gamma^\nu(\vec{p}_o, s_o; \vec{p}_o + \vec{k}_o, s') +$$

$$\Gamma^\nu(\vec{p}_o, s_o; \vec{p}_o - \vec{k}_f, s') e_\nu(\vec{k}_o) \frac{2m \wp_+(\vec{p}_o - \vec{k}_f)}{m^2 - (\vec{p}_o - \vec{k}_f)^2} e_\mu^*(\vec{k}_f) \Gamma^{\mu *}(\vec{p}_f, s_f; \vec{p}_o - \vec{k}_f, s') \Big)$$

5. 8

This amplitude can immediately be used to provide an input for a unitary scattering that includes pair annihilation into 2 quanta using the identifications described elsewhere[3]. The annihilation amplitude can be directly obtained in this case by replacing the appropriate particle four-momentum with an antiparticle four-momentum of reversed sign.

### B. Small coupling limit

Examining the (on-shell) form of the vertex function from equation 3. 11 and 3. 14, we see that

$$\langle \vec{p}', s' | \Delta h | \vec{k}_\gamma, \lambda; \vec{p}, s \rangle \stackrel{\bullet}{=} \delta^3(\underline{k}_\gamma + \underline{p} - \underline{p}') \frac{\Gamma^\mu(\vec{p}', s'; \vec{p}, s)}{2\pi} e_\mu(\vec{k}_\gamma, \lambda)$$

5. 9

Since the intermediate state is NOT a boundary state, the vertex function should not be evaluated using a boundary state for the intermediate state, and must be extracted from the full t matrix. However, in the weak coupling limit, one can obtain an approximate form for this factor by using a zeroth order form for the intermediate state:

$$\Gamma^\mu(\vec{p}', s'; \vec{p}, s) \approx \frac{q}{c} \overline{\mathbf{u}}(\vec{p}', s') \gamma^\mu \mathbf{u}(\vec{p}, s)$$

5. 10

The Dirac spinors have the following normalization properties

$$\overline{\mathbf{u}}(\vec{p},s)\mathbf{u}(\vec{p},s') = \delta_{s,s'}$$

$$\sum_s \mathbf{u}(\vec{p},s)\overline{\mathbf{u}}(\vec{p},s) = \wp_+(\vec{p}) = \left(\frac{\mathbf{p}+m\mathbf{1}}{2m}\right)$$

5. 11

We can immediately evaluate the spin sums using these expressions, and the property of the positive energy projector $\wp_+^2 = \wp_+$. This gives a form for the weak coupling amplitude given by

$$A(\vec{k}_f,\lambda_f;\vec{p}_f,s_f | \vec{k}_o,\lambda_o;\vec{p}_o,s_o) \approx$$

$$\left(\frac{1}{2\pi i}\right)^2 \left( \overline{\mathbf{u}}(\vec{p}_f,s_f)\frac{q}{c}\gamma^\mu e_\mu^*(\vec{k}_f,\lambda_f) \frac{\{m\mathbf{1}+\mathbf{p}_o+\mathbf{k}_o\}}{m^2-(\vec{p}_o+\vec{k}_o)^2-i0^+} e_\nu(\vec{k}_o,\lambda_o)\frac{q}{c}\gamma^\nu \mathbf{u}(\vec{p}_o,s_o) + \right.$$

$$\left. \overline{\mathbf{u}}(\vec{p}_f,s_f)\frac{q}{c}\gamma^\nu e_\nu(\vec{k}_o,\lambda_o)\frac{\{m\mathbf{1}+\mathbf{p}_o-\mathbf{k}_f\}}{m^2-(\vec{p}_o-\vec{k}_f)^2-i0^+} e_\mu^*(\vec{k}_f,\lambda_f)\frac{q}{c}\gamma^\mu \mathbf{u}(\vec{p}_o,s_o) \right)$$

5. 12

One can directly compare this with the QED result in second order:

$$A(\vec{k}_f,\lambda_f;\vec{p}_f,s_f | \vec{k}_o,\lambda_o;\vec{p}_o,s_o) \approx i^2 \left(\frac{e^2}{(2\pi)^2}\right)\overline{\mathbf{u}}(\vec{p}_f,s_f)\left\{\frac{\not{e}^*(\vec{k}_f,\lambda_f)\{m\mathbf{1}+\mathbf{p}_o+\mathbf{k}_o\}\not{e}(\vec{k}_o,\lambda_o)}{m^2-(\vec{p}_o+\vec{k}_o)^2} + \right.$$

$$\left. \frac{\not{e}(\vec{k}_o,\lambda_o)\{m\mathbf{1}+\mathbf{p}_o-\mathbf{k}_f\}\not{e}^*(\vec{k}_f,\lambda_f)}{m^2-(\vec{p}_o-\vec{k}_f)^2}\right\}\mathbf{u}(\vec{p}_o,s_o)$$

These results are seen to directly correspond in the weak coupling limit.

# VI. CONCLUSION

In this paper we have demonstrated that a unitary quantum-particle scattering amplitude can be extracted from a source-particle-detector three-body system in which the source and detector are disentangled except through their interaction with the particle. The requirements on the source-particle and detector-particle two-body interactions which allow this extraction to be made are (a) that these two-body interactions be unitary in the appropriate two-body spaces, (b) that they individually be consistent, when disentangled, with the kinematics implied by the exchange of a single, on-shell quantum and (c) that we can change the off-energy shell parameter and the kinematics of these two-body interactions to access a primary singularity at the quantum mass. We emphasize that this extraction is made possible by the fact that our Lorentz-invariant Faddeev-type formalism is *cluster decomposable.* In a previous communication[3] we have presented a model containing a quantum mass parameter which includes zero as a possibility, and in that case reproduces both quantum Coulomb scattering with the correct essential singularity in energy and the Bohr bound state spectrum in the non-relativistic limit. We also show that this model has an essential singularity at the mass of the quantum in the particle-antiparticle transition amplitude.

      Our choice of Compton scattering as the first physical process containing photons as boundary states that we calculate using our finite particle number formalism has turned out to be fortunate in a number of ways. The initial reason for the choice was simply that at some point we had to reproduce a standard QED result in the appropriate limit before most theorists would take our approach seriously, and that long ago a colleague of ours had challenged us to do this for Compton scattering. As we have explained above, by adopting an appropriate (Wheeler-Feynman) point of view[8,11] we soon discovered that the apparently difficult problem of modeling sources and detectors of photons could be solved kinematically, insuring the generality of the result. The second, but more systematically planned, factor that lead to success was that we had already solved the general problem of how to introduce anti-particles into our formalism[3]. This allowed us to use the conventional interpretation of negative energy particles as positive energy antiparticles to construct the usual causal propagator for the problem at hand. An unexpected, but in hindsight obvious, discovery to which our care with the discussion of source and detector led to was that the unit term in the S-matrix for the process whose T-matrix is the amplitude computed here corresponds physically to the self consistent calibration of the source and detector with the flux normalization of the transition amplitude. The fact that the factors we compute by our rules which clothe the three separate terms are identical, and hence cancel out independent of the structure of the source and detector, thus gives graphic proof that our unitary formalism does guarantee appropriately unitary results for extracted amplitudes.

We hope that by now we will have convinced some readers that we can in principle, and to some extent in practice, deal with the same problems that renormalized QED was created to handle, but in a non-perturbative way. We also hope that a few of our readers will themselves want to try to apply our non-perturbative methods to the more challenging problems raised by quark and gluon confinement, electro-weak unification, ….. If so, this paper will have served its purpose.

**Acknowledgements**:

J.V.L. would like to acknowledge the hospitality of the Stanford Linear Accelerator Center during the period when this research was conducted. The authors wish to thank Walter Lamb for careful reading and discussion of the manuscript, and Alex Markevich for a discussion. H.P.N. wishes to thank the director of SLAC for arranging the visitor program which enabled J.V.L. and H.P.N. to conduct this collaborative research.